\documentstyle[11pt,psfig]{article}
\twocolumn
 1

\def\etal{{et al.~}}

\def\HH{${\rm {H_2}}\,\,$}

\def\gs{\mathrel{\raise1.16pt\hbox{$>$}\kern-7.0pt
\lower3.06pt\hbox{{$\scriptstyle \sim$}}}}
\def\ls{\mathrel{\raise1.16pt\hbox{$<$}\kern-7.0pt
\lower3.06pt\hbox{{$\scriptstyle \sim$}}}}
\def\gtsima{$\; \buildrel > \over \sim \;$}
\def\ltsima{$\; \buildrel < \over \sim \;$}
\def\prosima{$\; \buildrel \propto \over \sim \;$}
\def\gsim{\lower.5ex\hbox{\gtsima}}
\def\lsim{\lower.5ex\hbox{\ltsima}}
\def\simgt{\lower.5ex\hbox{\gtsima}}
\def\simlt{\lower.5ex\hbox{\ltsima}}
\def\simpr{\lower.5ex\hbox{\prosima}}

\def\pp{\noindent\parshape 2 0truecm 17truecm 2truecm 15truecm}
\def\rf#1;#2;#3;#4 {\par\pp#1, #2, #3, #4. \par}

\def\pr{\ref@jnl{Phys.Rev}}

\def\ie{{\frenchspacing\it i.e. }}

\def\href#1;#2 {{\bf #1} : {\em #2}}

\def\beq#1{\begin{equation}\label{#1}}
\def\eeq{\end{equation}}
\def\beqa#1{\begin{eqnarray}\label{#1}}
\def\eeqa{\end{eqnarray}}

\def\HH{H$_2$ }
\def\H2p{H$_2^+$ }

\def\mH2p{H_2^+}

\title{DETECTING POP~III OBJECTS WITH NGST} 

\author{\bf A. Ferrara  $^{1,2}$, S. Marri $^{3}$\\
$^{1}$ Joint Institute for Laboratory Astrophysics, Boulder, USA\\
$^{2}$ Osservatorio Astrofisico di Arcetri, Firenze, Italy\\
$^{3}$ Dipartimento di Astronomia, Universit\'a di Firenze, Italy.}   
\date{}

\begin{document}
\maketitle
\centerline{ABSTRACT}
We discuss some aspects concerning the formation and the impact of
the first luminous structures in the universe (PopIIIs), 
with particular emphasis on their feedback effects on subsequent 
galaxy formation.
We argue that supernovae in these objects might provide, if detected, 
important constraints for cosmological models. Some of these constraints come
from the gravitational magnification of the supernova flux due to the
intervening cosmological matter distribution which results in different
predictions, for example, on rates, magnification probability,
detection limits and number counts of distant supernovae. We show that
NGST can tremendously contribute to such a study due to its advanced
technological capabilities. 


\centerline{1. POP III OBJECTS}

Current models of cosmic structure formation based on CDM scenarios
predict that the first collapsed, luminous (PopIIIs) objects 
should form at redshift $z\approx 30$ and have a total mass $M \approx 
10^6 M_\odot$ or baryonic mass $ M_b \approx 10^5 M_\odot$
(Couchman \& Rees 1986, Haiman \etal 1997, Tegmark \etal 1997). 
This conclusion is reached by requiring that the cooling time, $t_c$, 
of the gas is shorter than the Hubble time, $t_H$, at the formation epoch. 
In a plasma of primordial composition the only efficient
coolant in the temperature range $T\le 10^4$~K, the typical virial temperature
of PopIII dark matter halos, is represented by H$_2$ molecules whose abundance
increases from its initial post-recombination relic value to higher values 
during the halo collapse phase. It is therefore crucial to determine the 
cosmic evolution of such species in the early universe to clarify if small 
structures can continue to collapse according to the postulated hierarchical 
structure growth or if, lacking a cooling source, the mass build-up sequence 
comes to a temporary halt.

\centerline{1.1 Negative Feedback ?}

The appearance of PopIIIs is now thought to cause a partial
destruction of the available molecular hydrogen either in the intergalactic 
medium (IGM) and/or in collapsing structures; the result is a negative feedback 
on galaxy formation. This effect has been pointed out by HRL,
and works as follows.
As stars form in the very first generation of objects, the emitted 
photons in the energy band 11.2-13.6 eV are able to penetrate the gas  and 
photodissociate H$_2$ molecules both in the IGM and in the nearest collapsing 
structures, if they can propagate that far from their source. This negative 
feedback and its possible limitations are discussed by Ciardi, Ferrara \& Abel
(1998). In brief, these authors have studied the evolution of ionization and
dissociation spheres produced by and surrounding PopIIIs,
which are supposed to have total masses $M \approx 10^{6} M_{\odot}$
and to turn on their radiation field at a redshift $z=30$.
By a detailed numerical modelling of non-equilibrium radiative transfer,
they conclude that the typical size of the dissociated region is $R_d 
\approx 1-5 $ kpc. As the mean distance between PopIIIs at $z \approx 
20-30$ in a CDM model is $d \approx 0.01-0.1$ Mpc, \ie larger than $R_{d}$.
Thus, at high redshift, the photodissociated regions are not large enough 
to overlap. In the same redshift range, the soft-UV
background in the Lyman-Werner bands (photons leading to \HH 
photodissociation) when the intergalactic H and H$_{2}$ opacity 
is included, is found to be $J_{LW} \approx 10^{-28}-10^{-26}$ erg 
cm$^{-2}$ s$^{-1}$ Hz$^{-1}$. This value
is well below the threshold required for the negative
feedback of PopIIIs on the subsequent galaxy formation
to be effective before redshift $\approx 20$.   

\centerline{1.2 Positive Feedback}

\begin{figure}
\centerline{\psfig{figure=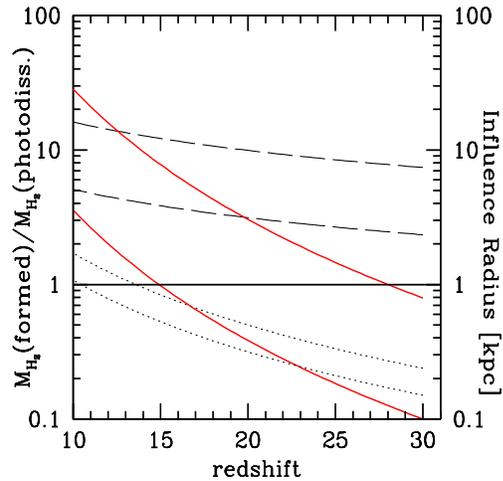,width=10.cm,height=11cm}}
\caption{{\em\label{fig1}Ratio between the $H_2$ mass formed
and destroyed by a PopIII as a function of the
multi-SN explosion redshift (solid lines); values larger
than one for the ratio define the epochs where PopIIIs
have a {\it positive feedback} on galaxy formation. Also shown are
the shell (proper) radius at cooling, $R_s(t_c)$ (dotted), and the
photodissociation (proper) radius, $R_d$ (dashed). The upper set of curves
refers to objects of mass $M=10^6 M_\odot$, whereas the bottom one corresponds
to larger objects, $M=10^7 M_\odot$. 
The cosmological parameters are $\Omega_b=0.05$, $h=1$.
}}
\end{figure}

In addition to the uncertain impact of negative feedback, Ferrara (1998)
has suggested that instead PopIIIs might be able to provide a  
{\it positive} feedback based on supernova (SN) explosions, which, 
under many aspects, is reminiscent of a scaled version of the explosive 
galaxy formation scenario introduced by Ostriker \& Cowie (1981) and put 
forward by many others. PopIIIs are very fragile due to their low
mass and 
shallow gravitational potential: only a few SNe are sufficient to blow-away 
(Ciardi \& Ferrara 1997)
their baryonic content and drive an expanding blastwave into the IGM,
which eventually becomes radiative and allows the swept gas to cool in a dense
shell. 
Typically a $H_2$ fraction $f=6\times 10^{-3}f_6$ is formed after
explosive events leading to the blow-away of PopIIIs.
In order to evaluate the impact of PopIIIs on the subsequent galaxy formation,
largely regulated by the availability of $H_2$ in this mass
(and redshift) range, it is useful to compare the $H_2$
mass production vs. destruction.
The ratio between the $H_2$ mass produced and destroyed
by PopIIIs is
\begin{equation}
\label{m+m-}
{M^+_{H_2}(z)\over M^-_{H_2}(z)}= \left({f\over f_{IGM}}\right)
\left[{R_s(t_c)\over R_d}\right]^3,
\end{equation}
where $R_s(t_c)$ is the radius of the SN-driven shell at one gas 
cooling time.
The post-recombination relic fraction of intergalactic $H_2$
is estimated to be $f_{IGM} \approx 2\times 10^{-6} h^{-1}$
(Palla \etal 1995; Anninos \& Norman 1996; Lepp, private communication).
The previous relation is graphically displayed in Fig. \ref{fig1},
along with the values of $R_s(t_c)$ and $R_d$. From that plot
we see that objects of total mass $M_6=1$ produce more $H_2$
than they destroy for $z\simlt 25$; larger objects ($M_6=10$)
provide a similar positive feedback only for $z\simlt 15$,
since they are characterized by a higher $R_d/R_s(t_c)$ ratio.
However, since in a hierarchical model larger masses form
later, even for these objects the overall effect should be
a net $H_2$ production.
This occurrence suggests that these first objects
might have a positive feedback on galaxy formation.

\centerline{2. LENSING OF HIGH$-z$ SNe}

\begin{figure}
\begin{center}
\centerline{\psfig{figure=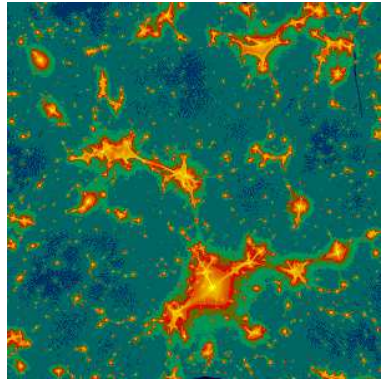,height=5cm}}
\end{center}
\caption{{\em\label{fig2} (Color Fig.) Magnification map for the 
LCDM model at redshift $z=8$; the strongest caustics (white regions)
correspond to magnifications $\mu\approx 30$.
}}
\end{figure}

If PopIIIs can be able to escape destruction of their \HH molecular content,
enough cooling will be available to initiate their gravitational collapse.
As the collapse proceeds, the gas density increases and stars are
likely to be formed. However, the final product of such star formation
activity is presently quite unknown. This uncertainty largely depends on
our persisting ignorance on the fragmentation process and its relationship 
with the thermodynamical conditions of the gas,
both at high $z$ and, in a less severe
manner, in present day galaxies. Ultimately, this prevents firm conclusions
on the mass spectrum of the formed stars or their IMF. This problem was already
clear more than two decades ago, as pioneering works (Silk 1977; Kashlinsky
\& Rees 1983; Palla, Salpeter \& Stahler 1983; Carr \etal 1984) could not reach
similar conclusions on the typical mass range of newly formed stars in
the first protogalactic objects. 
If the IMF is more of the standard type,
questions arise about its shape and median value. A common claim in the
literature is that the absence of metals (as it is the case in a collapsing
PopIII) should shift the peak and the median of the IMF toward higher masses.
Obviously, assessing the presence of massive stars that produce ionizing photons
and die as Type II supernovae would be of primary importance to clarify 
the role of PopIIIs in the reionization, photodissociation,  and 
rehating of the universe, and, in general, for galaxy formation.
Thus, studying the formation of the first stars has become one of the
most challenging problems in physical cosmology.
If SNe are allowed to occur at these very
high redshifts, they could outshine their host protogalaxy by orders of
magnitude and likely become the most distant observable sources since the
QSO redshift distribution has an apparent cutoff beyond $z\approx 4$.
Nevertheless, even for (future) large telescopes as NGST, VLT and Keck
this will not be an easy task, this requiring to reach limiting magnitudes
$\approx 32$ in the near IR to observe a SN exploding at $z=10$.
 
Marri \& Ferrara (1998, MF) have investigated the gravitational lensing 
(GL) magnification effects on high-$z$ SNe in three different cosmological
models: SCDM (Standard Cold Dark Matter:
$\Omega_M=1$), LCDM (Lambda Cold Dark Matter: $\Omega_M=0.4,
\Omega_\Lambda=0.6$), CHDM (Cold Hot Dark Matter: $\Omega_M=1
\Omega_\nu=0.3$); all models have $h=0.65$, and are normalized to the
present abundance of clusters. The main output of the ray-shooting
numerical simulations are a set of
magnification maps displaying the magnification, $\mu$  (\ie the source flux
enhancement factor) of a point source
located at a given spatial position inside the considered $4' \times 4'$
field of view. As an example we show in Fig. \ref{fig2} the magnification 
map relative to the LCDM model at redshit $z=8$, where the caustics are 
clearly identified; 
the maps for all models at different redshifts can be found in MF.
There are several unambiguous differences among the
three families of cosmological models that we can identify by analysing
the magnification maps. SCDM models give intense, but not very
numerous  caustics ($\mu \approx 30-40$) for $z_s \approx 3$; intermediate
magnification caustics appear at higher $z_s$ due to lower mass objects that
have not yet hierarchically merged into larger ones. LCDM models produce
the most intense caustics ($\mu \approx 50$), but most of the map is covered
with more diffuse and lower $\mu$ magnification patterns. Finally, CHDM models,
which form large structures later than $z=3$, show very rare intense events
and many moderate $\mu$ caustics. 
The magnification probability function presents a moderate degree of evolution
up to $z\approx 5$ (CHDM) and $z\approx 7$ (SCDM/LCDM). All models predict that
statistically large magnifications, $\mu \simgt
20$ are achievable, with a probability of the order of a fraction of percent,
the SCDM model being the most efficient magnifier.
All cosmologies predict that above $z\approx 4$ there is a
10\% chance to get magnifications larger than 3.
 
\centerline{3. NGST EXPECTATIONS} 

The advances in technology are making available a new generation of
instruments, some already at work and some in an advanced design phase,
which will dramatically increase our observational capabilities.
As representative of such class, we will focus on a particular instrument,
namely the Next Generation Space Telescope (NGST). In the following we
will try to quantify the expectations for the detection of high $z$ SNe
and the role that the gravitational lensing can play in such a search.

We will assume that NGST (i) is optimized
to detect radiation in the wavelength range from $\lambda_{min}=1\mu$m to
$\lambda_{max}=5\mu$m (\ie J-M bands), and (ii) can observe to a
limiting flux of ${\cal F}_{NGST}=10$~nJy in $10^{2}$ s in that range,
which should allow for low-resolution spectroscopic follow-up.
This can be achieved, for a
8-m (10-m) mirror size and a S/N=5, in about $2.6\times 10^4$~s
($1.1\times 10^4$~s)\footnote{Result obtained using
the NGST Exposure Time Calculator}
We also assume that a Type II SN has a black-body spectrum (Kirshner 1990)
with temperature
$T_{SN}$ and we fix its luminosity $L_{SN}=3\times 10^{42}$ (Woosley \&
Weaver 1986; Patat \etal 1994).
This constant luminosity plateau lasts for about $\approx 80 (1+z)$~days,
after which the object fades away.

\centerline{3.1 Apparent Magnitudes}

\begin{figure}
\leavevmode
\centerline{\psfig{figure=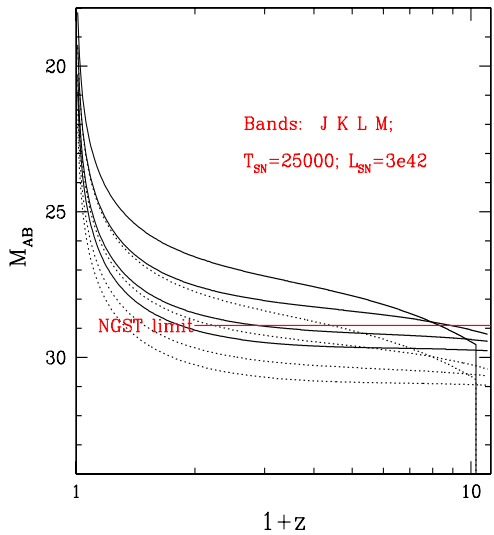,width=8.0cm,height=9cm}}
\caption{{\em\label{fig3} Apparent AB magnitude of a SN as a function
of
its explosion redshift in the four wavelength bands J, K, L, M  (from the
uppermost to the lowermost curve) for the SCDM model.
The case for a moderate ($\mu=3$) magnification (solid
curves) is compared to the one in which magnification is neglected (dotted).
The NGST flux limit is also shown; the vertical line
at high redshift in the J band is due to intergalactic absorption.
 }}
\end{figure}

Fig. \ref{fig3} shows the apparent AB magnitude of a SN as a function of
its explosion redshift in four wavelength bands (J, K, L, M) in the 
assumed NGST sensitivity range, and we compare it with the instrument 
flux limit.
Note that we have taken into account absorption by the IGM
at frequencies higher than the hydrogen Ly$\alpha$ line.
The plot also assumes that $T_{SN}=25000$~K, a value corresponding to the
temperature approximately appropriate to the first $15 (1+z)$~days after
the explosion (Woosley \& Weaver 1986). 
Including GL magnification
enhances dramatically the observational capabilities. In fact, as one can
see from the solid line set of curves in Fig. \ref{fig3}, even allowing
for a magnification $\mu =3$ only, which in all models has a magnification
probability larger than 10\% at high $z$, this pushes
the maximum redshift at which SNe can be detected up to $z \approx 9$

\centerline{3.2 Detection Limits}

\begin{figure}
\leavevmode
\centerline{\psfig{figure=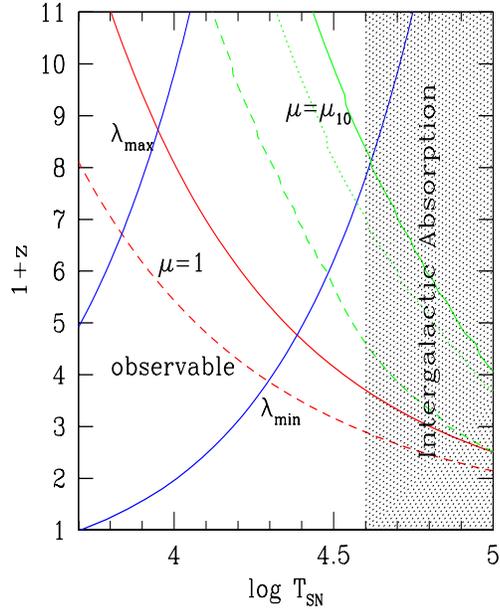,width=8.0cm,height=10cm}}
\caption{{\em\label{fig4} SN detection area (marked as "observable") in the
redshift-SN temperature, $T_{SN}$, plane neglecting GL magnification (curves
$\mu=1$) and with magnification $\mu_{10}(z)$ for the three cosmological
models: SDCM (solid curve), LCDM (dashed), CHDM (dotted). Note that for $\mu=1$
the SCDM and CHDM curves overlap. The area is also
bound by the lower and upper NGST wavelength limits, $\lambda_{min}$ and
$\lambda_{max}$, respectively. The shaded region shows the effect of
intergalactic absorption; the SN luminosity is $3\times 10^{42}$~erg~s$^{-1}$.
 }}
\end{figure}

A different way to appreciate the GL effects is illustrated by Fig. \ref{fig4}.
There we allow for the SN temperature to vary in the large interval $T_{SN}
\approx 4000 - 10^5$~K and we ask in which region of the  $(1+z)-T_{SN}$
parameter space NGST will be able to detect SNe in primordial objects.
This region is constrained by the requirement that the radiation from the
SN (or the maximum of its black-body spectrum) falls in between
$\lambda_{min}$ and $\lambda_{max}$ with a flux larger than ${\cal F}_{NGST}$.
In the absence of magnification, the upper redshift boundary of the detection 
area
(marked as "observable" in the Figure) is $4 < z< 8$ for $10^4~{\rm K}\simlt
T_{SN} \simlt 3\times 10^4~{\rm K}$ for SCDM and CHDM models (both having
$\Omega_\Lambda=0$) and $3 < z< 5.5$ for $8000~{\rm K}\simlt
T_{SN} \simlt 2.5\times 10^4~{\rm K}$ for LCDM. This difference is due to the
larger luminosity distance in LCDM models. We can now compare this result with 
the one obtained considering the GL magnification of the SN. 
As an indication, we use for $\mu$ the value of 
$\mu_{10}(z)$, \ie the largest magnification with a probability 
$P(\mu)> 10$\%: this quantity should represent a good compromise
between the two, usually conflicting, requirements that a high 
magnification and high
probability event occurs. In this case, the upper redshift boundaries (the three
uppermost declining curves in Fig. \ref{fig4}) of the
detection area are shifted towards higher redshift, in a way that depends
on the cosmological models, which are now clearly differentiated because of
the concomitant effect of their different $\mu_{10}(z)$ and luminosity 
distances.
Also, the total area is generally increased, thus considerably enhancing the
detection chances. The maximum redshift at which a SN can be observed
is now above $z=10$, the limit of our GL simulations, for all models, and it
can be extrapolated to $z \approx 12$ for SCDM.
From Fig. \ref{fig4}, it appears that, for a SN search aimed study,
the most effective improvement with respect to the planned NGST 
characteristics would be an extension of the bandpass into the mid-IR,
which would greatly enhance the maximum detection redshift of these
objects. 

\centerline{3.3 Number Counts}

Recently, Marri, Ferrara \& Pozzetti (1998), have derived the expected 
differential number counts, $N(m)$, of Type II SNe in the above
mentioned cosmological models, taking into account the effects of 
gravitational magnification. In brief, they calculate the expected
SN rate per unit (comoving) volume $\gamma(z)$ from semi-analytical 
models. Next, they calculate how magnification by gravitational lensing 
affects the evolution of the observed SN luminosity function (LF) 
(assuming that the local LF is the one suggested by Van  
den Bergh \& McClure 1994) using the following prescriptions:  
{\it i}) calculate the total number of SNe by integrating $\gamma(z)$
in a given redshift interval $\delta z$ around $z_i$
and in a certain cosmic solid angle $\omega$;
({\it ii}) assign a peak luminosity to each SNe by randomly sampling
the van den Bergh \& McClure luminosity function;
({\it iii}) assign to each SN a certain value of the
magnification, $\mu$, by randomly selecting a cell of the magnification
map relative to the appropriate
redshift interval and cosmological model. Magnification maps and
magnification probability distributions, $P(\mu)$, are those 
derived by MF.

\begin{figure}
\leavevmode
\centerline{\psfig{figure=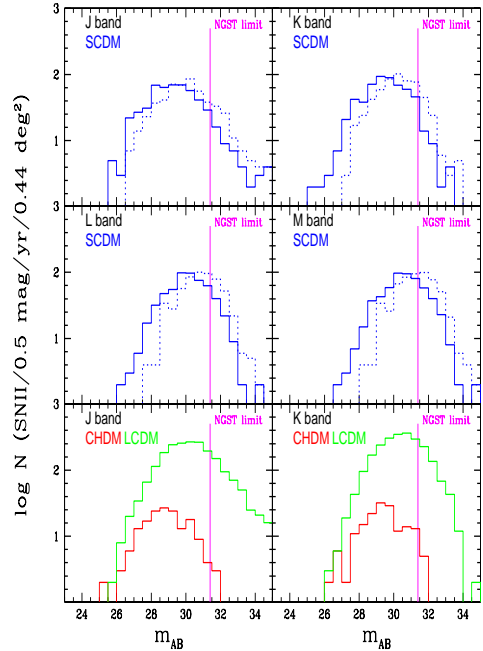,width=7.0cm,height=10cm}}
\caption{{\em\label{fig5}Differential number counts for the three 
cosmological models
considered as a function of apparent AB magnitude in $J$,$K$,$L$,$M$ bands
(LCDM and CHDM: $J$ and $K$ only). Also shown is the NGST limiting magnitude.
{\it Solid} ({\it dashed}) curves include (neglect) lensing magnification.
}}
\end{figure}

Following MF, $\delta z=0.2$; the value of $\omega$ is
taken as the sum of 100 NGST fields, \ie $0.44$~deg$^2$;
we suppose that these fields are surveyed for one year.
This experiment is well within the scheduled observational plans
and capabilities of NGST (Stockman \etal 1998).
Fig. \ref{fig5} shows the log of the differential SNe counts
as a function of AB magnitude.
The four upper panels contain the curves for the SCDM model and for
$J$, $K$, $L$, and $M$ bands, both including or neglecting
the effects of gravitational lensing on the LF.
The gravitational flux magnification allows the detection,
at a given limiting flux,  of a larger number of SNe.  For comparison,
we plot the NGST magnitude limit $AB=31.4$ (vertical line).
Thus, NGST should be able to reach the peak of expected SNe count
distribution, located at $AB\approx 29-30$
(depending on the model),
with roughly 1 mag gain brought by gravitational magnification.
The three cosmological models (SCDM, LCDM, CHDM)
predict a total number of (574, 2373, 164) SNe/yr in 100 surveyed
fields of NGST.

From these results it appears clear that by pushing SNe searches to
faint magnitudes, and exploiting the gravitational lensing magnification,
we will be able to build a statistically significant sample, which
will allow to test various cosmological models and
study the cosmic star formation history of the universe at epochs
otherwise difficult to investigate.

\vskip 1truecm
We are grateful to our collaborators in the project, B. Ciardi and L. Pozzetti 
for allowing presentation of results in advance of publication. 

\centerline{REFERENCES}

Anninos, P., \& Norman, M. L. 1996, ApJ, 460, 556

Ciardi, B.,  \& Ferrara, A. 1997, ApJ, 483, 5 

Ciardi, B.,  Ferrara, A. \& Abel, T. 1998, ApJ, submitted 

Carr, B. J., Bond, J. R. \& Arnett, W. D. 1984, ApJ, 277, 445

Couchman, H. M. P. \& Rees, M. J. 1986, MNRAS, 221, 53


Ferrara, A. 1998, ApJ, 499, L17

Haiman, Z., Rees, M. J., \& Loeb, A. 1997, ApJ, 476, 458

Kashlinsky, A. \& Rees, M. J. 1983, 205, 955

Kirshner, R. P. 1990, in "Supernovae", ed. A. G. Petschek, 
(Springer: New York), 59

Marri, S. \& Ferrara, A. 1998, ApJ, in press (astro-ph/9806053)

Marri, S., Ferrara, A. \& Pozzetti, L. 1998, ApJ, submitted 


Ostriker, J. P. \& Cowie, L. L. 1981, ApJL, 243, 127

Palla, F., Salpeter, E. E. \& Stahler, S. W. 1983, ApJ, 271, 632

Palla, F., Galli, D. \& Silk, J. 1995, ApJ, 451, 44

Silk, J. 1977, ApJ, 211, 638

Stockman, H. S., Stiavelli, M., Im, M., \& Mather, J. C. 1998, Science with
the Next Generation Space Telescope, ed. E. Smith \& A. Koratkar (ASP Conf. Ser.
), in press

Tegmark, M., Silk, J., Rees, M.J., Blanchard, A., Abel, T. 
\& Palla, F.  1997, ApJ, 474, 1    

van den Bergh S. \&  McClure R. 1994, ApJ, 425, 205

Woosley, S. E. \& Weaver, T. A. 1986, ARA\&A, 24, 205

\vfill
\end{document}